\documentclass[11pt]{article}
\usepackage{graphicx}
\begin{document}
 
\title{{\bf Distant globular clusters with anomalously small masses}}
\author{{\bf T.V. Borkova, V.A. Marsakov}\\
Institute of Physics, Rostov State University,\\
194, Stachki street, Rostov-on-Don, Russia, 344090\\
e-mail: borkova@ip.rsu.ru, marsakov@ip.rsu.ru}
\date{received July 29, 2002 \ accepted October 13, 2002}
\maketitle

\begin {abstract}
We found that 10 metal-poor globular clusters are greately distinguished
for  anomalously small masses on the "destruction rate--mass" plain.
As it turned out, these poor clusters, situated  15 kpc farther from the
Galactic centre,
are somewhat younger than the bulk of the metal-poor globulars, and have
anomalously red horizontal branches. All these clusters are supposed to belong
to the "young halo" subsystem, i.e. they are supposed to have been captured
by the Galaxy at different stages of their evolution. We discovered a
significant  correlation between the ages found from isochrones and masses of
globulars which lie at galactocentric
distances greater than the radius of the solar orbit. At the same time,
deficiency of distant massive clusters is noticeable with increasing distance
from the galactic centre. So we see unticorrelation between
the galactocentric distance and masses of distant clusters. Both relations
are negligible for the inner clusters of the Galaxy.
We assume that favourable conditions for violent dissipation with considerable
loss of mass are realised inside the protoglobulars which formed far from
the Galactic centre.
\end {abstract}

Van den Berg (1998, 2000) payed attention to the fact that all known to him
"young" globular clusters  with ages by $\approx3$ billion years
younger than the oldest clusters of the same metallicity lie at
galactocentric distances $R>15$\,kpc and have masses smaller than the average.
That is why he assumed that the initial mass spectra of globular
clusters formed in the outer halo of the Galaxy displace with time
towards smaller masses. Independently, we (Borkova, Marsakov, 2000) have found
that the average mass of clusters located within the radius of the
solar circle
somewhat increases, while in the outer halo it decreases noticeably with
growing galactocentric distance. The clusters of the outer halo have
markedly smaller masses than those of the thick disk, even among the youngest
objects (see Fig.\,12 and Table 2 in the papers by Borkova and Marsakov, 2000).
In the present paper this phenomenon is discussed in more details and an
attempt to explain it is made.

 As initial data we have used the compiled
catalogue of the measured characteristics of the globular clusters of the Galaxy
(Harris, 1996) and the summary catalogue of basic parameters for 145
clusters compiled on its basis by Borkova and Marsakov (2000) with
involvement of other sources. For the present paper we have recalculated
the parameters on the basis of a new version of the catalogue by Harris (1996)
compiled for 1999 June 22, where the distances and luminosities of clusters
of the inner region of the Galaxy were mostly revised. We found the masses
from integral stellar magnitudes of clusters using a constant
mass-to-light ratio $M/L_V=3$, where the mass $M$ and luminosity $L_V$
are expressed in solar
units. It should be borne in mind, however, that the constancy of the
ratio $M/L_V$ is not satisfied in all the cases. In particular, the
drop in luminosity of clusters with a given mass will be observed in the
case of poor population of their red giant branch. Similarly, the mass of a
cluster distant from the galactic centre may be underestimated due to a considerable fraction of small-mass stars preserved from tidal
destruction. Fig.\,1a shows a relationship
between the galactocentric distance and the mass of clusters. The
dashed line restricts the region of slow evolution of globular clusters the
existence of which is theoretically grounded in the paper by Surdin and
Arkhipova (1998). The upper line corresponds to the critical value of mass
resistent to the effects of dynamical friction that leads to deceleration
of a massive stellar cluster moving through the field stars and to its
destruction at the centre of the  Galaxy under the action of tidal forces,
while the lower one corresponds to the effects of dissipation from tidal
"shocks" when the cluster flies through the galactic plane. It can be
stated with high probability that the clusters lying on the diagram outside
of this region are at the end of their lifetime. The solid lines in Fig.\,1a
represent straight regressions for the regions inside and outside the
solar circle ($R_{\odot}=8.5$\,kpc). The corresponding correlation 
coefficients
are shown in the figures. It can be seen that in the inner region the
inspected data do not reveal variation of the average mass of clusters with
increasing distance from the galactic centre. However, in the outer region of
the Galaxy the observed anticorrelation is different from zero beyond the
errors ($r=0.3\pm0.1$). The application of the $t$ criterion has shown that
correlation between the mass and the galactocentric distance of globular
clusters in the outer halo takes place at a confidence level of about
95\,\% despite the small value of correlation coefficient. The slope
of the regression does not decrease, and the correlation coefficient remains
beyond the errors even if six the most remote points are excluded from the
diagram. According to Surdin and Arkhipova (1998), such distant clusters
did not undergo a considerable  dissipation and dynamical friction.
This is why the initial mass distribution in the outer halo remained almost
unaffected.
Thus, a question arises why the deficiency of massive globular clusters
increases in the outer halo with increasing galactocentric distance
(the lower right  angle in the diagram is practically empty).

The present-day theory of dynamical evolution of globular clusters offers a
possibility of estimating the mass loss rate of a large number of clusters.
In particular,  Gnedin and Ostriker (1997) calculated the destruction rate
of 119 clusters of galaxies caused by the combined action of
two--body relaxation, tidal destruction, and shocks with the disk and
bulge. The authors disregarded the  energy increase of the clusters
caused by their interaction with giant molecular clouds, following the
conclusions made by Chernoff et al. (1986). The tangential components of
velocities of each cluster, which were lacking for the calculation of the galactic
orbits were derived by Gnedin and Ostriker (1997)  statistically from the
kinematic model for the galactic globular cluster  system. In the recently
published paper by Dinesku et al. (1999) the destruction rate for 38
globular clusters with the measured tangential and radial components of space
velocities were computed. On the whole, the authors note a good agreement with
the results of Gnedin and Ostriker (1997). That is why, in order to use as
many clusters as possible, we considered  to make use of the
calculation results  of destruction rates derived by Gnedin and
Ostriker (1997). Fig.\,1b presents the relation between the mass of clusters
and the total destruction rates  $\nu$. The light symbols denote
the clusters situated  at a distance of $R>15$\,kpc from the galactic
centre. It can be seen that a sufficiently reliable decrease
of the average mass with increasing destruction rate is observed. However,
rather a narrow sequence consisting of 10 distant clusters parallel to the
main one separates sharply. Their masses are approximately an order of
magnitude smaller than the average masses of clusters at the same values of $\nu$.
The 10 clusters are shown in Fig.\,1 by open triangles. Only one cluster
with an anomalously small mass (Pal 3) fell in the sample of Dinesku
et al. (1999), and in the diagram destruction rate --- mass, plotted
from the data of this paper, it lies lower than the main relation by the same
value.

It is interesting that three of the isolated clusters of anomalously small mass
are usually believed to belong  to the Sagittarius dwarf spheroidal galaxy. These are
the clusters Arp\,2, Ter\,7 and Pal\,12. They are closely spaced and located
at a distance of about 18\,kpc from the centre of the Galaxy. Apart from
them, there are located another two clusters from the same galaxy, Ter\,8 and
NGC\,6715. They did not fall into our group of clusters of anomalously
small mass. Although the mass of Ter\,8 is the same as that of the three
isolated clusters, but its destruction rate has not been found.
The mass of the cluster NGC\,6715 is  two orders of magnitude larger.
A group of five clusters with $R\approx100$\,kpc
and $log(M/M_{\odot})\approx4.5$ is distinguished in Fig.\,1, but in fact,
they are spaced by
 a few dozen kiloparsec. Such a distant, but massive cluster NGC\,2419
has a very low central density which is more characteristic of small-mass
clusters, ($log(\rho)=1.5$), with an average density logarithm for all
clusters of the Galaxy approximately equal to three.
For the clusters of anomalously small mass that we have selected this value is
$\sim0.2$.

Check the ages of the clusters included in this group and  the
existence of clusters with masses above the average among the "young" clusters
in the entire sample. It is of interest to see if there is any time variation
of the average mass of the clusters inside and outside the solar
circle
separately. For this purpose, we make use of the ages from the paper by
Borkova, Marsakov (2000) where, based on the ages of 47  sources taken from
literature
and 336 individual determinations reduced to the unified scale, the weighted
mean estimates of ages for 63 globular clusters are computed. In averaging,
a two-stage iteration procedure with assigning  weight both to each of the
used source of ages of the clusters and to each individual determination of
individual age. The internal accuracy of the obtained estimates is equal to
$\pm0.9$ billion years. In the present paper, in addition to the original
list, we have computed the ages of another 14 clusters. The diagrams
$t-log(M/M_{\odot})$ are displayed in Fig.\,2. Mass  variation as a function of age is not observed in the inner region of the Galaxy. The
correlation coefficient within the errors is equal to zero, see the figures in
Fig.\,2a. In the outer region the correlation turns out to be rather high
($r=0.6\pm0.1$), it remains significant ($r=0.4\pm0.1$) even if four the
youngest clusters are disregarded (i.e. the points far away from the
centre of distribution). It should be noted that for the most frequently used
method, when the ages of globular clusters are estimated from the luminosity
of the stars of the turnoff points, the difference in ages between the clusters
with different content of heavy elements depends strongly on the adopted relation
between luminosity and metallicity of the horizontal branch stars. This is
why, the younger age of comparatively metal-rich globular clusters is still in
question. The method of estimation of ages from the effective temperatures of
the turnoff point stars, which assumes coincidence of the positions of the red
giant branches of clusters of the same metallicity, can be used to obtain
only the difference in ages between the clusters being compared.
In this case, the relative ages prove to be more reliable. The ages of the
clusters of anomalously small mass distinguished in Fig.\,1b have been obtained
by both methods in about ten and a half papers. And both the methods
suggest their age to be smaller as compared with the majority of metal-poor
clusters.

The presence of an explicit correlation between mass and age
in the outer halo suggests a time shift of the initial mass spectrum  of
globular clusters towards small masses. Indeed, massive enough clusters in
Fig.\,2b appear only at $t\sim 12$ billion years. However, attention should be
 drawn
to the fact that in the range of ages (12--14 billion years) clusters are
observed in a wide range of masses and all the clusters of this age with
relatively great masses (NGC\,362, 1851, 3201, 6981, IC\,4499, and Rup\,106)
lie in the range of galactocentric distances ($9\div 18$) kpc, whereas small-mass
clusters (Pal 3, 4, 5, 13, 14, Arp\,2, and Pyxis) are situated at far
greater distances ($18\div 102$) kpc. It means that at the same age the decisive
factor that causes the small mass of clusters is their distance from
the galactic centre.

So, the decreasing of the average mass of globular clusters with diminishing
age is observed only in the outer regions of the Galaxy. The inner
clusters do not reveal such an effect. Note that 10 distant clusters of
anomalously small mass selected by us proved to be younger and with an average
metallicity $<[Fe/H]>=-1.36\pm0.14$ possess anomalously red horizontal
branches (except for Arp\,2) non-typical for globular clusters
genetically related with the Galaxy. All of them belong to the subsystem of
the "young halo", i.e. they were assumingly captured by the Galaxy at
different  stages of their evolution (see Zinn (1993), Da Costa and Armandroff
(1995)).

It looks as if only far from the centre of  the Galaxy, conditions are
created that lead to formation of globular clusters with anomalously small
masses.
The farther from the centre of the Galaxy the more frequently these
conditions are realized: almost all the (five out of six) clusters with $R>60$\,kpc
have masses several times smaller than the average over the Galaxy. Probably,
here the predicted by Agekian (1962) process of violent dissipation of
globular clusters turns to be efficient, as a result of which they may
lose the greater part of their masses during a short lifetime.
It is caused  by the fact that at the stage of
condensation of a massive diffuse cloud of a protocluster, every star formed
inside it tends to its centre due to the cloud attraction. When passing
by the centre of inertia of the
cluster, part of the stars have time to accelerate so much that they leave it.
In this case the percentage of stars "evaporated" from the cluster depends
on the degree of original sphericity of the protocluster. Only the protoclusters,
formed at a considerable distance from any massive objects, including those
formed at a considerable distance from the Galaxy centre, can get
a "regular" shape. The existence in the outer halo of older clusters quite
rich in stars can be explained in this case by the fact that these clusters
(located, mainly, much closer to the galactic  centre than the younger
ones) formed under the condition of close interaction with other massive
objects and at the stage of the protocloud could not acquire a regular
spherical shape. (The  known very distant massive cluster NGC\,2419 is not
only an older one, but its central density, as has already been noted, more
corresponds to a  density of low-mass clusters.)
In contrast to them, younger clusters were captured by the Galaxy from a
relatively poorly populated, more distant  region of the intergalactic
space, where they lost a considerable part of mass at the initial stage
of evolution. Within such an explanation, the deficiency of massive clusters
at large galactocentric distances becomes clear. This explanation is somewhat
hindered by the large range of metallicities of our clusters of
anomalously small mass ($-1.8<[Fe/H]<-0.6$). However, two clusters with
$[Fe/H]>1.0$ belong to Saggitarius and one more, the youngest one having at the
same time the smallest mass, Pal\,1, can be referred with the same degree of
probability both to globular clusters and open clusters (Van den Berg, 2000).
The remaining six small-mass clusters have metallicities close to
$[Fe/H]\approx -1.6$ with $\sigma_{[Fe/H]}\approx 0.1$.

Certainly, each of the correlations discussed here turns out
to be of little evidence because of the small number of the observed clusters and
large errors in determination of their parameters. Nevertheless, since
the distances, masses, ages, radial velocities and morphological composition
of the horizontal branches of clusters are determined independently, the
assembly of all the results suggests that in the outer halo  an
isolated group of mass globular clusters really exists. To draw a reliable conclusion
if the deficiency of small-mass young globular clusters at the Galaxy periphery
is related to the initial mass function variation of these objects with the age,
or to the existence of violent dissipation in their evolution, it is
necessary to measure proper motions of distant clusters. The elements of
galactic orbits computed on their basis make it possible to define the
supposed sites of their formation and estimate reliably the destruction
rate of every cluster.

{\bf Acknowledgements}: 
The work was done under the support of the RFBR through grants 00--02--17689
and 02--02--06911.


\begin{figure*}
\centering
\includegraphics[angle=-90,width=17cm]{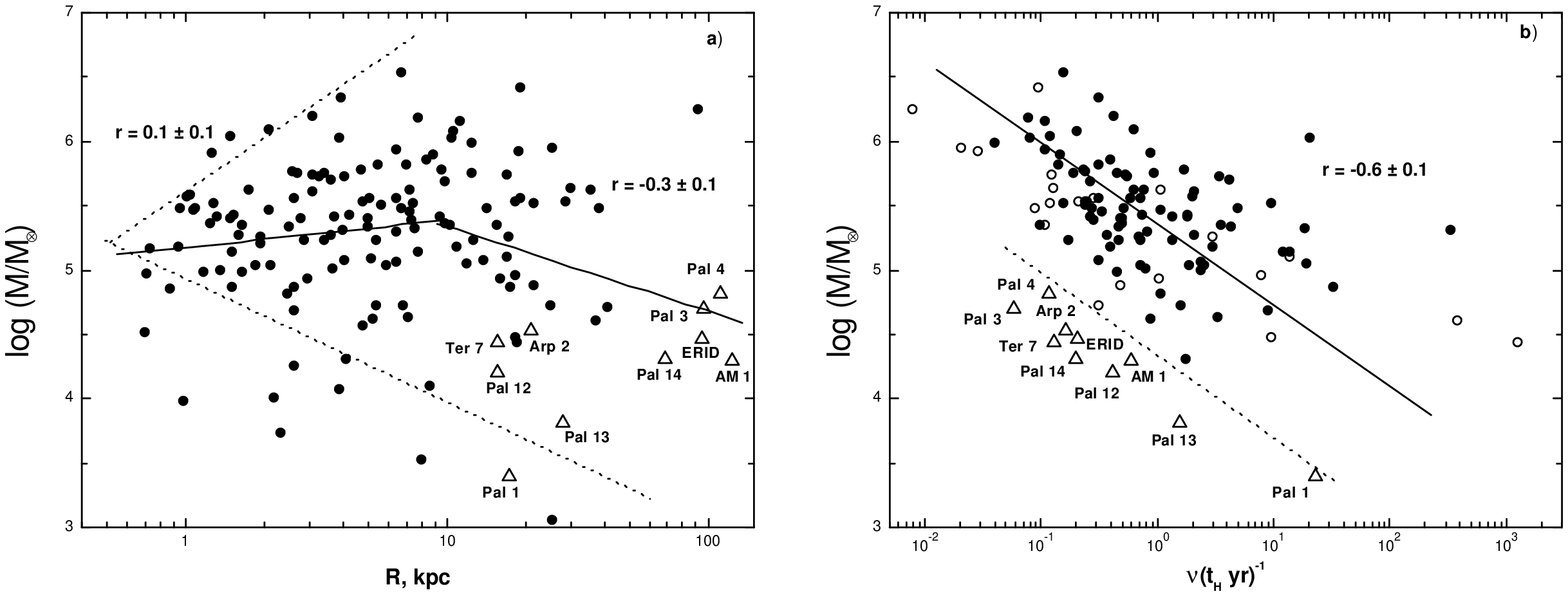}
\caption{The relationship of the mass and the observed galactocentric
distance R --- (a), and that of the mass and the total dissipation
velocity $\nu$ taken from Gnedin,
Ostriker (1997) --- (b). The dotted line in diagram (a) shows the "cone of
survival" according to the paper by Surdin and Arkhipova (1998). The thin
lines are the stright lines of the regression in the regions inside and
outside the solar orbit radius, while $r$ are the corresponding correlation
coefficients. The open triangles are for the clusters lying below the dotted
line in diagram (b). The values of $\nu$ in diagram (b) are expressed in
units reverse to Hubble time adopted for convenience to be equal to 10 billion
years. The open symbols (circles and triangles) denote the clusters lying
farther than 15 kpc from the centre of the Galaxy. The solid line is a
regression stright line for the main sequence of clusters, and the dotted
line parallel to it separates the sequence of clusters of anomalously small
mass shown by open triangles. Their names are beside them.}
\label{fig1}
\end{figure*}

\newpage

\begin{figure*}
\centering
\includegraphics[angle=-90,width=17cm]{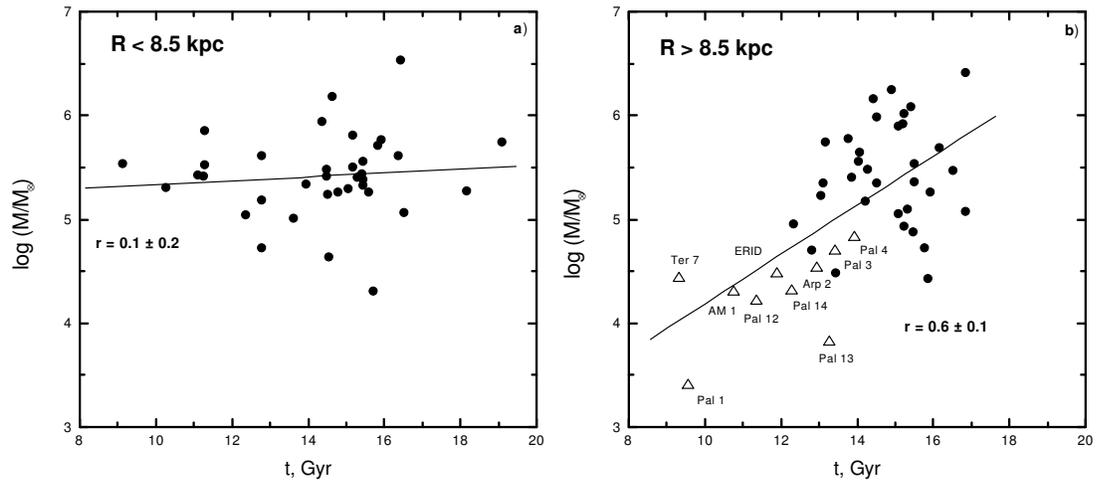}

\caption{The mass--age relationship for the clusters lying inside (a) and
outside (b) the solar circle. The stright lines are root-mean-square regressions,
$r$ are the corresponding correlation coefficients. The open triangles in
diagram (b) show the clusters of anomalous mass the same as in Fig.\,1b.}
\label{fig2}
\end{figure*}

\end{document}